\documentclass[utf8]{frontiersSCNS} 

\usepackage{url,hyperref,lineno,microtype,subcaption}
\usepackage[onehalfspacing]{setspace}
\usepackage{epsfig}
\usepackage{hyperref}
\usepackage{aas_macros}
\usepackage[T1]{fontenc}

\def\keyFont{\fontsize{8}{11}\helveticabold }
\def\firstAuthorLast{Young} 
\def\Authors{Peter R. Young\,$^{1,2,*}$}
\def\Address{$^{1}$NASA Goddard Space Flight Center, Greenbelt, MD 20771, USA \\
$^{2}$Northumbria University, Newcastle upon Tyne, NE1 8ST, UK  }
\def\corrAuthor{NASA Goddard Space Flight Center, Code 671, Greenbelt, MD 20771, USA}

\def\corrEmail{peter.r.young@nasa.gov}

\def\ion#1#2{#1\,{\sc #2}}
\newcommand{\as}{${^\prime}{^\prime}$}
\newcommand{\lam}{$\lambda$}

\begin{document}
\onecolumn
\firstpage{1}

\title[EUV and Soft X-ray Solar Spectroscopy Missions]{Future Prospects for Solar EUV and Soft X-ray Spectroscopy Missions} 

\author[\firstAuthorLast ]{\Authors} 
\address{} 
\correspondance{} 

\extraAuth{}

\maketitle

\begin{abstract}

Future prospects for solar spectroscopy missions operating in the extreme ultraviolet (EUV) and soft X-ray (SXR) wavelength ranges, 1.2--1600~\AA, are discussed. NASA is the major funder of Solar Physics missions, and brief summaries of the opportunities for mission development under NASA are given. Upcoming major solar missions from other nations are also described. The methods of observing the Sun in the two wavelength ranges are summarized with a discussion of spectrometer types, imaging techniques and detector options. The major spectral features in the EUV and SXR regions are identified, and then the upcoming instruments and concepts are summarized. The instruments range from large spectrometers on dedicated missions, to tiny, low-cost CubeSats launched through rideshare opportunities.

\tiny
 \keyFont{ \section{Keywords:} keyword, keyword, keyword, keyword, keyword, keyword, keyword, keyword} 
\end{abstract}

\section{Introduction}

This paper discusses prospects for future extreme ultraviolet (EUV) and soft X-ray (SXR) spectroscopic missions observing the Sun. It is based on presentations given by the author at the 2018 American Geophysical Union meeting and the 2019 RHESSI Workshop.\footnote{Presentations available at \url{https://pyoung.org/talks}.}

The EUV and SXR contain a huge number of narrow emission lines from many different elements and ion species formed in all layers of the solar atmosphere, and 
spectroscopy plays a critical role in understanding the physics occurring there. Doppler shifts yield true plasma velocity measurements, while broadening of the lines beyond the known thermal and instrumental broadening can yield information about turbulence, wave propagation or non-equilibrium processes. Modeling of the atomic processes occurring in the atmosphere is necessary to uncover emission line ratio diagnostics for individual ions that can yield the plasma temperature and density. The CHIANTI atomic database \citep{2016JPhB...49g4009Y} is widely used in Solar Physics for modeling emission lines. Combining lines from multiple ions and elements allows the differential emission measure---a quantity that describes the amount of plasma at different temperatures---to be obtained, and the ratios of element abundances. The latter is important for understanding the so-called ``FIP effect": a phenomenon in the corona and solar wind whereby elements with low first ionization potentials (FIPs) are found enhanced over those with high FIPs \citep{2015LRSP...12....2L}. A detailed review of EUV and SXR spectroscopic diagnostics   is given by \citet{2018LRSP...15....5D}.

In addition to emission lines, the EUV and SXR regions contain continuum emission from \emph{bremsstrahlung} (free-free) and radiative recombination (free-bound). The former is a continuous function that dominates at high energies, while the latter shows jumps in the emission corresponding to ionization edges of the emitting species. A difficulty in measuring the continuum is the subtraction of the detector background level which may not be well-characterized. At X-ray wavelengths the continuum can be important for estimating the plasma temperature, particularly for low-resolution data. Since the continuum is dominated by emission arising from collisions with hydrogen, the intensity is strongly dependent on the hydrogen abundance. Hence line-to-continuum ratios can yield ``absolute'' element abundance ratios (i.e., relative to hydrogen), whereas line spectra alone usually yield only relative abundance ratios \citep{2014ApJ...786L...2W}. Continuum emission has been demonstrated to be a significant contributor to radiative losses during flares, e.g., \citet{2012ApJ...748L..14M}.

The article is organized as follows. Since most solar missions are funded by NASA either through NASA-led projects, or as contributions to non-US missions then  Sect.~\ref{sect.nasa}  gives details about the Heliophysics program at NASA and the mission opportunities that are available. Sect.~\ref{sect.foreign} presents upcoming major solar missions from other nations. Sect.~\ref{sect.wvl} defines the wavelength regions used in this article, and Sect.~\ref{sect.instr} summarizes the observing strategies that are usually employed in the EUV and SXR regions. Properties of the EUV spectrum are given in Sect.~\ref{sect.euv-spec}, and the new EUV instruments and concepts are described in Sect.~\ref{sect.euv}. The SXR spectrum and instruments are described in Sects.~\ref{sect.sxr-spec} and \ref{sect.sxr}, and a final summary is given in Sect.~\ref{sect.summary}.

Many acronyms are used in this article for the names of instruments and missions and so a summary is given in Table~\ref{tbl.acronyms}.

\begin{table}[t]
\caption{Mission and instrument acronyms used in this work. The launch year in brackets is given where appropriate.}
\begin{center}
\small
\begin{tabular}{lp{3.5in}ccc}
\noalign{\hrule}
\noalign{\smallskip}
\noalign{\hrule}
\noalign{\smallskip}
Acronym & Name & Launch & Mission & Type$^1$ \\
\noalign{\hrule}
\noalign{\smallskip}
AIA & Atmospheric Imaging Assembly & 2010 & SDO & MI \\
CDS & Coronal Diagnostic Spectrometer & 1995 & SOHO & MI\\
ChemiX & Chemical composition in X-rays & -- & Interhelioprobe & MI\\
COSIE & COronal Spectrographic Imager in the Extreme ultraviolet & -- & -- & C \\
CubIXSS & CubeSat Imaging X-ray Solar Spectrometer & -- & -- & Cu\\
DAXSS & Dual Aperture X-ray Solar Spectrometer & -- & INSPIRESat-1 & Cu \\
EIS  & EUV Imaging Spectrometer & 2006 & \emph{Hinode} & MI \\
EIT  & EUV Imaging Telescope & 1995 & SOHO & MI\\
ESIS & EUV Snapshot Imaging Spectrograph & 2019 & -- & R\\
EUI & Extreme Ultraviolet Imager & 2020 & Solar Orbiter & MI\\
EUVI & Extreme Ultraviolet Imager & 2006 & STEREO & MI \\
EUVST & Extreme Ultraviolet High-Throughput Spectroscopic Telescope & -- & Solar-C & MI\\
EUNIS & Extreme Ultraviolet Normal Incidence Spectrograph & 2006,07,13 & -- & R\\
EVE & Extreme ultraviolet Variability Experiment & 2010 & SDO & MI\\
FOXSI & Focusing Optics X-ray Solar Imager & 2012,14,18 & -- & R\\
FURST & Full-Sun Ultraviolet Rocket SpecTrograph & -- & -- & R\\
Hi-C & High-resolution Coronal Imager &2012,18 & -- & R \\
HRTS & High Resolution Telescope and Spectrograph & 1975--1992 & -- & R \\
IRIS & Interface Region Imaging Spectrograph & 2014 & -- & M \\
MaGIXS & Marshall Grazing Incidence X-ray Spectrometer & -- & -- & R\\
MinXSS & Miniature X-ray Solar Spectrometer & 2016,18 & -- & Cu\\
MUSE & MUlti-slit Solar Explorer & -- & -- & C\\
NIXT & Normal Incidence X-ray Telescope & 1989--93 & -- & R\\
RDS & Rotating Drum x-ray Spectrometer & -- & KORTES & MI\\
RHESSI & Reuven Ramaty High Energy Solar Spectroscopic Imager & 2002 & -- & M\\
SDO & Solar Dynamics Observatory & 2010 & -- & M\\
SERTS & Solar EUV Rocket Telescope and Spectrograph & 1989--99 & -- & R \\
SMM & Solar Maximum Mission & 1980 & -- & M \\
SNIFS & Solar eruptioN Integral Field Spectrograph & -- & -- & R\\
SOHO & Solar and Heliospheric Observatory &1995 &-- & M\\
SPICE & Spectral Imaging of the Coronal Environment & 2020 & Solar Orbiter & MI\\
STEREO & Solar Terrestrial Relations Observatory & 2006& -- & M \\
STIX & Spectrometer / Telescope for Imaging X-rays & 2020 & Solar Orbiter & MI\\
SUMER & Solar Ultraviolet Measurements of Emitted Radiation  & 1995 & SOHO & MI\\
SunCET & Sun Coronal Ejection Tracker & -- & -- & Cu\\
SUVI & Solar Ultraviolet Imager & 2016,18 & GOES & MI\\
SWAP & Sun Watcher with Active Pixel System detector and Image Processing & 2009 & Proba-2 & MI \\
TRACE & Transition Region and Coronal Explorer &1998 & -- & M\\
UVCS & Ultraviolet Coronagraph Spectrometer & 1995 & SDO & MI\\
VERIS & VEry high Resolution Imaging Spectrograph &2013 & -- & R\\
VISORS & VIrtual Super-resolution Optics with Reconfigurable Swarms & -- & -- & Cu\\
XRT & X-Ray Telescope & 2006 & \emph{Hinode} & MI \\
\noalign{\hrule}
\noalign{\smallskip}
\multicolumn{5}{l}{$^1$ C--Concept; Cu--CubeSat; M--Mission; MI--Mission Instrument; R--Rocket}\\
\end{tabular}
\end{center}
\label{tbl.acronyms}
\end{table}

\section{NASA Heliophysics and Mission Opportunities}\label{sect.nasa}

Heliophysics is one of the four science sub-divisions within the NASA Science Mission Directorate (SMD), with the others being Astrophysics, Planetary Science and Earth Science. For financial year FY20, the total NASA budget was \$23B with SMD receiving \$7B. Heliophysics has the smallest budget of the four science divisions, receiving \$725M in FY20. Within Heliophysics, there are five main science areas: Solar Physics, Heliospheric Physics, Space Weather, Geospace Physics, and Ionospheric, Thermospheric and Magnetospheric (ITM) Physics. The present article is specifically about Solar Physics missions. New Heliophysics missions are launched with a frequency of 1--2 years, with the most recent being the Ionospheric Connection Explorer (ICON) in 2019, and the Parker Solar Probe (PSP) and Global-scale Observations of the Limb and Disk (GOLD) missions in 2018.

NASA missions broadly fall into four categories:
\begin{enumerate}
    \item Large, ``LWS-Class" strategic missions. These are strategic missions  funded under the Living with a Star (LWS) program and directed by NASA, with a Principal Investigator (PI) at a NASA center. Budgets are typically around \$1B and the most recent missions are SDO (launched 2010), the Van Allen Probes (2012) and Parker Solar Probe (PSP, 2018). The next mission will be the Geospace Dynamics Constellation (GDC).
    \item Solar-Terrestrial Probes (STP). These are also strategic missions directed by NASA, but the PI can be based outside of NASA. The budget is typically around \$500M and the most recent missions were \emph{Hinode} (launched 2006), STEREO (2006) and the Magnetospheric Multiscale Mission (2015). The next will be the Interstellar Mapping and Acceleration Probe (IMAP), to be launched in 2025.
    \item{Explorer missions} These are competitively-selected missions led by a PI. There is the Small Explorer program (SMEX) and the Medium Explorer program (MIDEX), with budgets of around \$100M and \$250M, respectively. Solar Physics explorers include RHESSI (launched 2002) and IRIS (2013), both SMEXes.        
    \item Mission of Opportunity (MOO). These are cheaper missions and can be selected in combination with a larger mission, as part of foreign partner mission, or as an International Space Station mission. Examples include EUVST and GOLD.
\end{enumerate}

The process for selecting PI-led missions and MOOs typically follows the pattern of a Call for Proposals, a review followed by a down-select, funded Phase~A studies for the advancing mission concepts, followed by a final selection. The time between proposal submission and final selection can be up to two years. The most recent SMEX selection was the Polarimeter to Unify the Corona and Heliosphere (PUNCH) in 2019 and the next
Call for Proposals is expected in the 2021--2022 time-frame. The most recent MIDEX Call for Proposals was released in 2019, with five missions selected for Phase-A studies in 2020. One of these, MUSE, is discussed in Sect.~\ref{sect.muse}.

In selecting the strategic LWS and STP missions, NASA takes guidance from the Decadal Survey of Solar and Space Physics performed by The National Academies of Science, Engineering and Medicine. The most recent  report\footnote{Available at \url{https://www.nationalacademies.org/our-work/a-decadal-strategy-for-solar-and-space-physics-heliophysics}.} was published in 2012, covering the 2013--2022 period, and the next is expected in 2023.


In addition to the above opportunities, NASA also has options for smaller-scale projects to be flown on sounding rockets and balloons. CubeSats and SmallSats are increasingly common and are typically launched through ``rideshare" arrangements to minimize costs. NASA also funds instruments to be placed on the International Space Station but this has not been used for solar EUV and SXR experiments so far.\footnote{The SolACES EUV irradiance spectrometer \citep{2014SoPh..289.1863S} was part of an ESA instrument package that was operational on the ISS from 2008 to 2017, and Sect.~\ref{sect.crystal} discusses the Russian  KORTES mission expected to be placed on the ISS in the near future.}


\section{Non-US solar missions}\label{sect.foreign}

The lack of large NASA missions in the current decade is compensated by three new missions from India, China and  Russia. Aditya-L1 is an Indian mission to be flown to the L1 Lagrange point with a launch date in 2022. The payload consists of a visible light coronagraph, an ultraviolet imaging telescope, two X-ray spectrometers (Sect.~\ref{sect.pin}) and three in situ instruments. 

China will be launching the Advanced Space-Based Solar Observatory \citep[ASO-S;][]{2019RAA....19..156G} into a Sun-synchronous orbit in 2022. The instruments to be flown are: a full-disk vector magnetograph,  Lyman-$\alpha$ imaging telescopes covering both the disk and corona, a full-disk white light telescope, and a hard X-ray imaging telescope. 

The Solar and Heliosphere Explorer---also known as Interhelioprobe---is a Russian mission consisting of two identical spacecraft expected to be launched in the 2026--27 timeframe \citep{2016Ge&Ae..56..781K}. The mission profile is similar to the European Space Agency's Solar Orbiter, with the spacecraft reaching as close as 0.3~AU to the Sun and up to 30$^\circ$ above the ecliptic plane. The payloads will include both remote sensing and in situ instruments. The former include EUV and SXR imaging telescopes, a coronagraph, a magnetograph and several high-energy instruments. The ChemiX SXR spectrometer is described in Sect.~\ref{sect.crystal}.

\section{X-ray and EUV wavelength regions}\label{sect.wvl}

For the purposes of this article, we take an unconventional definition of the soft X-ray (SXR) region to be between 1.2 and 50~\AA\, and the EUV to consist of two regions:  between 50--500~\AA\ (EUV-A) and 500--1600~\AA\ (EUV-B). We choose these divisions based on basic technology restrictions since this article is discussing space instrumentation.
Normal incidence optics are possible above 50~\AA, although with the restriction of using multilayer coatings below 500~\AA. The 1600~\AA\ limit is set by the fact that the longest-wavelength, strong transition region line is the \ion{C}{iv} line at 1550~\AA. 
The lower limit for soft X-rays is set at 10~keV (1.24~\AA) which is the usual definition taken by X-ray astronomers for the division between thermal and non-thermal plasma. A thermal plasma is one with Maxwellian particle distributions and is typically represented by one or two isothermal plasma components, or a continuous differential emission measure distribution. A non-thermal plasma shows enhanced emission above 10~keV beyond what is expected from a thermal electron bremsstrahlung spectrum, and is typically associated with accelerated electrons produced during a flare.


\section{Instrument options and constraints}\label{sect.instr}

In this section the different options for EUV and SXR spectrometers are discussed. Spatial resolution is necessary for measuring the morphology and evolution of most solar features so the options used in the EUV and SXR regions are discussed in Sect.~\ref{sect.imaging}. A wide range of detectors are used in these wavelength regions and  are briefly summarized in Sect.~\ref{sect.detector}.

Spatial resolution is generally given as an angular resolution in  arcseconds, with 1\as\ corresponding to 725~km on the Sun for an observatory at 1~AU. Spectral resolution is defined by the ratio of the wavelength of an emission line to its instrumental width, i.e., $\lambda/\Delta\lambda$, although for X-rays, wavelength is often substituted for energy and so resolution is given by $E/\Delta E$.
In general, higher spatial and spectral resolutions are easier to achieve at longer wavelengths. Thus IRIS has achieved 0.4\as\ spatial resolution and 50\,000 spectral resolution around 1300--1400~\AA, whereas EIS has achieved 3\as\ and 3000, respectively around 200~\AA. SXR crystal spectrometers can achieve spectral resolutions of $>1000$ but with no imaging capability.

\subsection{Spectrometer types}\label{sect.type}

The ultimate goal of solar observing at any wavelength is to simultaneously perform high resolution 2D imaging and spectroscopy---so-called \emph{integral field spectroscopy}. That is, each pixel in a 2D image will have its own high resolution spectrum with which to apply plasma diagnostics. At SXR wavelengths the pixels of a detector can have intrinsic energy resolution capability and options  are discussed in Sect.~\ref{sect.detector}.
For  EUV and longer wavelengths, the solution for simultaneous imaging and spectroscopy is to reconfigure the 2D image such that the individual ``pixels" can be processed through a standard slit--grating--detector set-up, and then re-constituted into an image by processing software. Sect.~\ref{sect.snifs} describes a new instrument concept that will use this approach. This type of integral field spectroscopy has been well-tested for ground-based astronomy telescopes using microlens arrays, fibre optic cables and image slicers \citep[e.g.,][]{2001PASP..113.1406L}. The problem at EUV wavelengths is that photon fluxes are low and the reconfiguration of the input image requires extra optical elements that would reduce the instrument throughput.

The next-best approach to integral field spectroscopy is \emph{imaging slit spectroscopy}, whereby the Sun is imaged through a narrow slit and the image is dispersed in the direction perpendicular to the slit with a grating.
This leads to a two-dimensional image on the detector that has wavelength in one direction and spatial information in the other. The first solar imaging slit spectrometer in the EUV was  HRTS, a rocket experiment flown several times between 1975 and 1992. The first spacecraft imaging slit spectrometers were CDS, SUMER and UVCS on SOHO. No slit spectrometers in the SXR region have yet been flown. The key disadvantage of slit spectrometers is that, to build up two-dimensional images, the slit has to be scanned step-by-step to build up a \emph{raster}. A solar active region typically has a size around 200\as\ and so a scan with a 1\as\ slit with 30~second exposure times would take 100~minutes. Because of this, a slit spectrometer is best used in tandem with a coronal imaging instrument such that the spectrometer observes a relatively small field-of-view and the evolution can be placed in context with the active region images.

\begin{figure}[t]
\centerline{\epsfxsize=5in\epsfbox{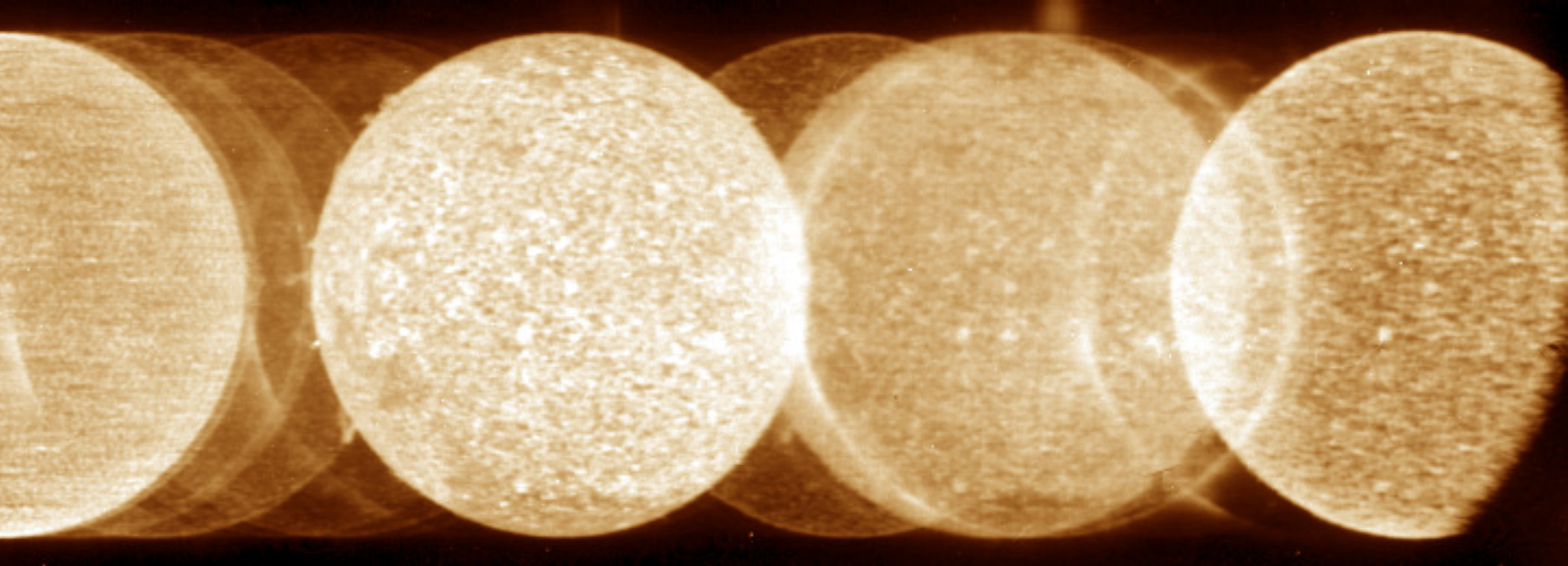}}
    \caption{A \emph{Skylab} S082A overlappogram from 1974 January 26 22:35~UT. Wavelength increases from the left to right. The bright half-Sun image at the left is from \ion{O}{iv} \lam554.4, and the bright partial image at the right is \ion{O}{v} \lam629.7. The brightest image to the left of center is from \ion{He}{i} \lam584.3.}
    \label{fig.skylab}
\end{figure}

Without a slit, an imaging spectrometer produces so-called ``overlappograms": each emission line in the spectrum will yield its own image of the Sun which then appear side-by-side in the detector image, spaced by the wavelength separations of the lines. This was most famously done with the S082A instrument on \emph{Skylab} in the 1970's (Figure~\ref{fig.skylab}). The CDS and EIS instruments were both flown with ``slot" options, i.e., slits that were significantly wider than the spectral resolutions of the instruments, thus producing rectangular images at the location of each line. For strong, relatively isolated lines the images are mostly pure and can be used for scientific analysis \citep[e.g.,][]{2009ApJ...695..642U}. As shown later in this article, slitless spectrometer designs are being considered for several mission concepts mainly to get around the limited field-of-view problems of slit spectrometers.

Spectrometers with no intrinsic spatial resolution are used for solar irradiance measurements and consequently require accurate radiometric calibration. Spectral resolution is usually low, but the EVE instrument \citep{2012SoPh..275..115W} on SDO achieved resolutions of 100 to 1000 in the 100--1000~\AA\ range, which has been sufficient to measure Doppler shifts in some circumstances \citep{2011SoPh..273...69H}. At SXR wavelengths, solar flares are usually the main focus and, since the emission largely comes from a single compact source, then valuable spectroscopic measurements can be made without the need for imaging. In recent years the development of very compact X-ray spectrometers means that solar SXR spectra can be obtained with low-cost missions, as described in Sect.~\ref{sect.pin}.

\subsection{Imaging options}\label{sect.imaging}

\begin{figure}[t]
    \centering
    \includegraphics[width=6in]{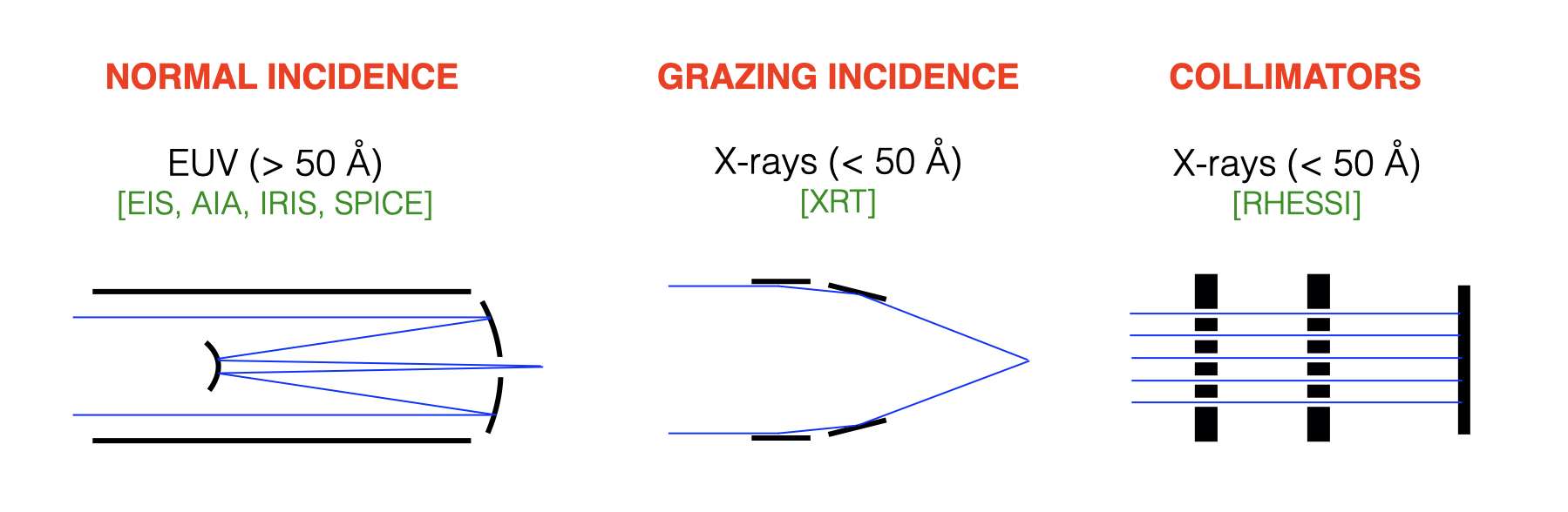}
    \caption{A simple guide to the imaging techniques employed in the EUV and SXR wavelength regions. Blue lines correspond to ray paths and black lines to optical elements}
    \label{fig.smiley}
\end{figure}

Figure~\ref{fig.smiley} illustrates the methods used to focus EUV and SXR radiation. The preferred method is to use normal incidence reflections from the mirror(s), but this is only possible above 50~\AA. Below this wavelength the incoming radiation is simply absorbed by the surface. Between 50 and 500~\AA\ normal incidence optics are only possible by applying multilayer coatings to the surfaces. These are alternating layers of a heavy and light element (Mb and Si are common choices) which lead to strongly enhanced reflectivity over a narrow wavelength range (typically 10--20\%\ of the central wavelength). The first solar telescopes to use multilayer coatings were flown on sounding rockets in the mid-to-late 1980's \citep{1987Sci...238...61U,1988Sci...241.1781W} and sub-arcsecond spatial resolution was demonstrated \citep{1990Natur.344..842G}. Multilayer coatings are ideal for solar imaging as they naturally yield  a narrow bandpass that can be tuned to specific emission lines in the EUV spectrum. The first multilayer imaging telescope flown on a spacecraft was EIT \citep{1995SoPh..162..291D} on SOHO, launched in 1995, followed by TRACE in 1998, EUVI on the STEREO spacecraft in 2006, SWAP on the Proba-2 spacecraft in 2009 \citep{2013SoPh..286...43S}, AIA on the SDO spacecraft in 2010 \citep{2012SoPh..275...17L}, SUVI on the GOES-16 and 17 spacecraft in 2016 and 2018 \citep{2019SPIE11180E..7PV}, and EUI on Solar Orbiter in 2020 \citep{2020A&A...642A...8R}. Use of multilayer coatings for spectroscopy was pioneered with the SERTS rocket program \citep{1993uxrs.conf..301D} and the first spacecraft spectrometer to use them was EIS on \emph{Hinode} \citep{2007SoPh..243...19C}. The best spatial resolution currently possible with normal incidence optics in the EUV and SXR is around 0.1\as, which is limited by the cost associated with the figuring of large mirrors. The Hi-C sounding rocket achieved 0.3\as\ \citep{2014SoPh..289.4393K}.

Above 500~\AA, broadband optical coatings can be used, giving high and relatively uniform sensitivity over a wide wavelength range, which is particularly valuable for spectroscopy. Examples include SUMER on SOHO and IRIS. 

Below 50~\AA\ direct imaging requires grazing incidence optics. As illustrated in Figure~\ref{fig.smiley}, the incoming radiation is focused off a pair of mirrors that are usually implemented in a cylindrical design (a Wolter-type telescope), with multiple concentric cylinders increasing the effective area. The XRT on \emph{Hinode} is an example of this design. The disadvantage of grazing incidence is that irregularities on the optical surfaces are magnified, leading to worse imaging performance than normal incidence optics. XRT achieved a 2\as\ resolution, which compares with 0.3\as\ for the Hi-C sounding rocket telescope \citep{2014SoPh..289.4393K}, which is currently the best performance for a solar EUV multilayer telescope.

Collimators are the least attractive option for imaging, but are often necessary at X-ray wavelengths. At their crudest they achieve spatial resolution simply through restricting the field-of-view.  A more sophisticated approach was used by RHESSI \citep{2002SoPh..210....3L}, which had two collimating grids. By rotating the spacecraft continuously every four seconds, a modulation pattern is built up that can be inverted to yield an image of the Sun. A spatial resolution up to 2\as\ can be achieved this way, but the method has limitations due to low dynamic range and a difficulty of resolving multiple sources. Note that a collimator approach is necessary in the hard X-ray wavelength region, hence grazing incidence optics were not an option for RHESSI.

A step beyond normal incidence optics is the use of Fresnel Zone Plates (FZPs) or Photon Sieves, which could yield spatial resolutions in the 10's of milliarcsecond range, i.e., an order of magnitude improvement over the best currently achieved in the EUV. The FZP is an idealized diffractive optic consisting of a circular plate with concentric rings cut into it. The placement of the rings is chosen to enable constructive interference of the transmitted light to yield a focused image. In practice, the rings are replaced with circles of dots or small slits---hence the name ``photon sieve"---and they were first discussed in terms of X-ray imaging by \citet{2001Natur.414..184K}. The key drawback with photon sieves is that the telescope needs a very long focal length of 100~m or more to achieve the highest spatial resolution. This requires a high-precision, formation-flying mission. For example, to yield sharp images, the transverse displacements of the optics spacecraft must be maintained to the size of a detector pixel, which is typically about 10~$\mu$m. This technology is becoming available now and, along with the photon sieve technology, is actively pursued by NASA \citep[e.g.,][]{2011SPIE.8148E..0OD,2018AcAau.153..349C}. In terms of spectroscopy, a photon sieve is of interest because an adjustment of the focal length can lead to a sampling of different parts of an emission line, thus revealing red-shifted or blue-shifted plasma. The first solar mission to employ a photon sieve will be VISORS, a CubeSat project funded by the NSF and led by the University of Illinois at Urbana-Champaign with a launch date in 2023. It will consist of two CubeSats flying in formation. One will host the detector and the other the photon sieve. The \ion{He}{ii} 304~\AA\ line is targeted and a spacecraft separation of 40~m coupled with a 75~mm sieve diameter will yield a spatial resolution of, at best, around 0.1\as.

\subsection{Detector options}\label{sect.detector}

CCDs are the standard sensors used for EUV and SXR imaging spectroscopy. They are usually used in a back-illuminated configuration and are directly sensitive to the incoming photons. At wavelengths above around 500~\AA, however, the sensitivity declines and it is common to use a microchannel plate to convert the photons to clouds of electrons. These can then be detected directly with an anode detector, such as for the SUMER and UVCS instruments on SOHO \citep{1995SoPh..162..189W,1995SoPh..162..313K}, or converted to visible photons via a phosphor screen and then detected with a CCD, which was done for CDS on SOHO \citep{1995SoPh..162..233H}.

The readout time of CCDs is slow. For example the 4k $\times$ 4k detectors of the AIA instrument on SDO take about 3 seconds. 
CMOS sensors (also referred to as Active Pixel Sensors, APS) have a much faster readout performance and are likely to be used more extensively in the coming decade. Like CCDs, they can be used in a back-illuminated configuration, as demonstrated by EUI on Solar Orbiter \citep{2020A&A...642A...8R}. A further advantage is that they are much more robust to harsh radiation environments, hence they have been used for the imaging instruments on board Solar Orbiter and PSP, which are operating far beyond the protective bubble of the Earth's magnetosphere.

Both CCD and CMOS detectors offer energy resolution in the SXR region, as the detectors are able to measure the numbers of electron-hole pairs created in the silicon by the incoming photons. (The Lyman-$\alpha$ line of \ion{Fe}{xxvi} produces almost 2000 pairs, for example.) The spectral resolution is limited to about 100~eV. Combining a CCD or CMOS with a focusing telescope enables imaging spectroscopy and examples are discussed in Sections~\ref{sect.phoenix}, \ref{sect.foxsi} and \ref{sect.ssaxi}.

The compact, silicon X-ray detectors from the company AMPTEK have proven very successful for obtaining moderate resolution, disk-averaged solar soft X-ray spectra from a package with dimensions approximately 15~mm. A detector was first flown on NASA's NEAR asteroid mission in 1996 in order to measure the solar X-ray flux. This measurement is often important for planetary science missions as solar X-rays illuminate the body's surface and the resulting fluorescent X-ray spectrum leads to valuable composition data. Similar detectors were also flown on SMART-1, Chandrayaan-1 (both lunar missions) and the Mercury MESSENGER mission. The solar data have proven valuable for solar flare studies \citep[e.g.][]{2015ApJ...803...67D}. The most recent planetary spacecraft with a solar X-ray spectrometer is the Indian Chandrayaan-2 mission \citep{2020SoPh..295..139M}, which was launched in 2019. 

The AMPTEK detectors  are either silicon PIN (Si-PIN) or silicon drift detectors (SDD) and they both come in an all-in-one configuration referred to as X-123. SDDs are more expensive but allow higher count rates and slightly better spectral resolution. Spectral coverage is typically 1--20~keV and resolution is about $E/\Delta E=30$ to 50, which is sufficient to resolve spectral features from which element abundances can be derived \citep{2015ApJ...803...67D,2018SoPh..293...21M}.

In terms of solar missions, the SphinX instrument \citep{2013SoPh..283..631G} on board the Russian CORONAS-Photon mission used an earlier version of the AMPTEK detector, while the X-123 was flown on the MinXSS-1 and 2 Cubesats \citep{2018SoPh..293...21M} in 2016 and 2018, although the latter failed before data could be obtained. New missions are described in Sect.~\ref{sect.pin}.

Microcalorimeter detectors would be a breakthrough for solar SXR spectroscopy, with two orders of magnitude improvement in energy resolution over silicon detectors. They can be arranged in an array format, thus yielding simultaneous 2D imaging and high-resolution spectra when combined with an imaging telescope.
Development has so far focussed on X-ray astronomy missions and the Japanese astrophysics spacecraft \emph{Hitomi}  was the first  to successfully operate  a microcalorimeter as part of the Soft X-ray Spectrometer (SXS).
Sadly \emph{Hitomi} perished after obtaining only a single observation of the Perseus galaxy cluster \citep{2016Natur.535..117H}. The SXS had a $6\times 6$ 2D pixel array with energy resolution of 7~eV between 0.3 and 12~keV \citep{2018JATIS...4b1402T}, which is comparable to crystal spectrometers (Sect.~\ref{sect.crystal}).

The high count rates from the Sun pose a problem for microcalorimeters compared to other astrophysical sources, and development progress lags behind that in Astrophysics. Some discussion of the scientific possibilities of microcalorimeters is given in the White Paper of \citet{2010arXiv1011.4052L}, while technical aspects are covered in \citet{2010SPIE.7732E..38B} and \citet{2013ITAS...2300705B}.

\section{The EUV spectrum}\label{sect.euv-spec}

\begin{figure}[t]
    \centering
    \includegraphics[width=\hsize]{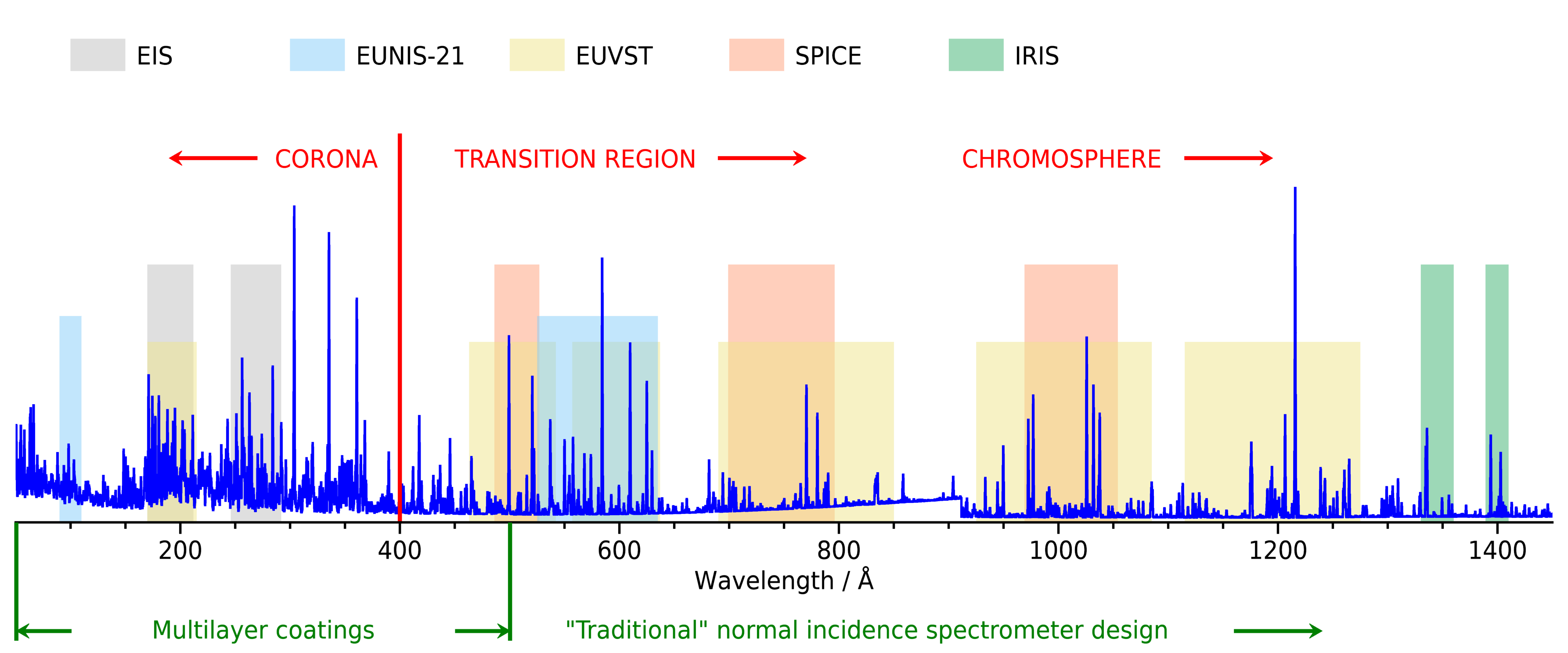}
    \caption{The blue line shows a synthetic spectrum generated with CHIANTI. The colored boxes denote the wavelength ranges of the five instruments listed at the top of the figure. The ``atomic dividing line" is denoted in red and the ``technology dividing lines" are shown in blue. See the main text for more details on these terms.}
    \label{fig.euv}
\end{figure}

Figure~\ref{fig.euv} shows a synthetic EUV spectrum computed with CHIANTI, with the wavelength ranges of five spectrometers indicated. 
An ``atomic dividing line" in the spectrum is found at about 400~\AA, below which most of the emission lines in the spectrum are formed in the corona corresponding to temperatures $>$~0.8~MK, and above which most lines are formed in the transition region and chromosphere, with temperatures $>20$~kK and $<0.7$~MK. This dividing line is close to the 500~\AA\ dividing line discussed earlier, below which multilayer coatings are required to enable normal incidence optics. Another important dividing line is 912~\AA, which is the Lyman limit corresponding to the hydrogen ionization edge. Although this limit does not impact technology limitations, it does mark a point below which there are no photospheric or low chromosphere lines in the spectrum. These lines are valuable as wavelength fiducial markers due to the small Doppler shifts in these regions.

Below 400~\AA, there are two key groups of iron lines. Between 170 and 212~\AA\ there are the strongest lines from the coronal iron ions \ion{Fe}{ix--xiv}, formed at temperatures 0.8--3.0~MK, which have been extensively studied with the \emph{Hinode}/EIS instrument and have also been targets for the EUV multilayer imaging telescopes mentioned in Sect.~\ref{sect.imaging}. Between 90 and 140~\AA\ there is group of strong transitions from \ion{Fe}{xviii--xxiii}, formed between 7 and 14~MK that offer excellent diagnostics of active regions and flares. The lines have been observed at low spectral resolution with EVE \citep{2012ApJ...755L..16M}, and \citet{2021arXiv210306156D} advocate for a future instrument to observe this region at higher resolution. 

Above 400~\AA\ the dominant transitions are the Lyman-$\alpha$ line of hydrogen, and the 584~\AA\ line of neutral helium. The most abundant elements of the lithium, beryllium, sodium and magnesium isoelectronic sequences all give rise to strong lines, which continue down below 400~\AA\ (Table~\ref{tbl.strong}).

\begin{table}[h]
\caption{Wavelengths (in \AA) of strong lines in the EUV.}
\begin{center}
\begin{tabular}{ccccc}
\noalign{\hrule}
\noalign{\smallskip}
\noalign{\hrule}
\noalign{\smallskip}
Element & Li-like & Be-like & Na-like & Mg-like \\
\noalign{\hrule}
\noalign{\smallskip}
C & 1548.2, 1550.7 & 977.0 & -- &--\\
N & 1238.8, 1242.8 & 765.2 & -- & -- \\
O & 1031.9, 1037.6 & 629.7 & -- &--\\
Ne & 770.4, 780.4 & 465.7 & -- & -- \\
Mg & 609.8, 624.9 & 368.1 & -- & -- \\
Si & 499.4, 520.7 & 303.3 & 1393.8, 1402.8 & 1206.5 \\
Fe & 192.0, 255.1 & 132.9 & 335.4, 360.8 & 284.2\\
\noalign{\hrule}
\end{tabular}
\end{center}
\label{tbl.strong}
\end{table}

Although the $>400$~\AA\ region mostly contains cool emission lines, there are an important set of coronal lines that are due to so-called ``forbidden'' transitions. That is, transitions that occur within the ground atomic configurations of the ions that have small but non-zero transition rates. Crucially these transitions enable spectrometers operating in the $>400$~\AA\ region to have some coronal capability and this has been exploited by the SPICE and IRIS spectrometers discussed in the next section. 

\section{EUV instrumentation}\label{sect.euv}

There are several EUV instruments currently operating, including imagers on STEREO, SDO,  GOES-16 and 17, and Solar Orbiter. In terms of spectroscopy, the key instruments are EIS, IRIS and SPICE which are all imaging slit spectrometers.
 EIS \citep{2007SoPh..243...19C} has been operating on the Japanese \emph{Hinode} mission for over 14 years. It obtains high resolution spectra in two bands at 170--212 and 246-292~\AA, with a spatial resolution of 3--4\as. Spectral resolution is 3000--4000, and the wavelength bands mostly include lines formed in the temperature range 0.5--20~MK.
Our definition of EUV for this article includes wavelengths below 1600~\AA\ and so  
IRIS \citep{2014SoPh..289.2733D}, with its two channels at 1332--1358 and 1389--1407~\AA, falls within our remit and was launched in 2013. Spatial resolution is 0.3--0.4\as, the best of any of the EUV/SXR spacecraft instruments observing the Sun, and spectral resolution is around 50\,000. The narrow wavelength bands give access to mostly chromospheric and transition region lines, with the exceptions of coronal forbidden lines of \ion{Fe}{xii} and \ion{Fe}{xxi} (formed at 1.5 and 11~MK, respectively), although the former is weak and rarely observed. IRIS also has a slitjaw imager, enabling the slit to be placed within simultaneous high-cadence chromospheric and/or transition region images.

SPICE \citep{2020A&A...642A..14S} was launched on the Solar Orbiter spacecraft in 2020 February and the optical design is similar to EIS but with two longer wavelength bands at 699--796 and 969--1054~\AA. The latter also has sensitivity to second order lines in the 485--527~\AA\ range. The spatial and spectral resolutions are 4\as\ and 1500--2000, respectively. Note that, since Solar Orbiter will reach distances of 0.3~AU from the Sun, then the effective spatial resolution will be up to a factor three better compared to a spectrometer at 1~AU. The wavelength bands give more complete temperature coverage than EIS and IRIS, particularly through the chromosphere and transition region, but the lines formed in the coronal range 1--4~MK are significantly weaker than those observed by EIS. Two strong forbidden lines of \ion{Fe}{xviii} and \ion{Fe}{xx} are formed at 7 and 10~MK, respectively, and should be important for active region and flare studies.

EVE is an irradiance spectrometer on board SDO that has two spectrographs with wavelength ranges 50--370 and 350--1050~\AA\ (MEGS-A and B, respectively). The former failed in 2014 but MEGS-B continues to operate.

The sections below discuss new mission concepts in the EUV wavelength range, arranged according to the spectrometer type (Sect.~\ref{sect.type}).

\subsection{Integral field spectrometers (SNIFS)}\label{sect.snifs}

The only integral field spectrometer for EUV wavelengths is SNIFS, which was selected for flight on a sounding rocket with a launch in 2024. It uses an array of  ``mirrorlets" to extract small portions of the input image and send them to a grating in such a manner that the final image on the detector appears as an array of 1D spectra, one for each mirror. To prevent the spectra from overlapping, the spectrum bandpass is restricted  and the grating is rotated relative to the mirrorlet array so that the dispersion axis is tilted relative to the detector axes. 
An earlier concept was described in \citet{2016JGRA..121.8250C}, where these principles are explained in greater detail.

SNIFS will use a $72\times 72$ mirrorlet array with each mirror corresponding to a spatial element of 0.45\as. Hydrogen Ly-$\alpha$ at 1216~\AA\ is the principal target and the nearby \ion{Si}{iii} 1206~\AA\ and \ion{O}{v} 1214~\AA\ lines will also be observed. Note that the optical elements of the system are a primary mirror, the mirrorlet array, a focusing grating and a detector; the mirrorlets replace the slit of a standard imaging slit spectrometer design.

\subsection{Imaging slit spectrometers}

\subsubsection{Solar-C\_EUVST}\label{sect.euvst}

Solar-A and Solar-B are Japanese missions that were renamed \emph{Yohkoh} and \emph{Hinode} on launch. The concept for the  successor Solar-C mission has evolved over time, but Japan and the USA have both recently approved the Solar-C\_EUVST concept \citep{2020SPIE11444E..0NS}, which is scheduled for launch in 2026. Based on the Large European module for solar Ultraviolet Research concept \citep[LEMUR;][]{2012ExA....34..273T} that was a component of a larger, multi-instrument configuration of Solar-C, EUVST is an imaging slit spectrometer with a slitjaw imager (the latter similar to that flown on IRIS).
The design is something of a hybrid of EIS and SPICE. The primary mirror and grating combine multilayer coatings and a broadband boron carbide coating to yield high efficiency in the longer UV wavebands of 690--850, 925--1085 and 1115--1275~\AA, together with the EUV waveband of 170--215~\AA. The two longer wavelength bands also have sensitivity to lines in the second spectral order, giving access to the wavelength ranges 463--542 and 557--637~\AA. Combined, these wavelength ranges are critical to giving EUVST the most complete temperature coverage of any solar spectrometer, with strong emission lines formed throughout the chromosphere, transition region, corona and flaring corona. The spatial resolution of 0.4\as\ is also comparable to IRIS, and EUVST will be the first spacecraft to achieve this resolution for wavelengths below 1000~\AA. 

\subsubsection{EUNIS}

EUNIS is a successor to the series of SERTS rocket experiments flown in the 1980's and 90's by a team at NASA Goddard Space Flight Center. It contains two independent imaging slit spectrometers with spatial resolutions of 3--4\as\ and spectral resolutions of 2000 to 2500,  and it has been flown in 2006, 2007 and 2013. The fourth flight has been delayed because of the COVID-19 pandemic but should take place in 2021. A new grating and multilayer coating will be used to obtain the first high resolution solar spectra in  the wavelength region 92--112~\AA\ for several decades. In addition to observing the strong \ion{Fe}{xviii} and \ion{Fe}{xix} emission lines in this region (solar activity permitting), the spectra will also be used to characterize cooler lines that contribute to the 94~\AA\ imaging channel of AIA. The second spectrometer channel covers 525 to 635~\AA.

\subsubsection{MUSE}\label{sect.muse}

MUSE \citep{2020ApJ...888....3D} is, at the time of writing, one of five mission concepts under review by NASA for selection as a MIDEX, with a launch date in the 2025--26 timeframe. It is a slit spectrometer, but with the novelty of having 37 closely-separated, parallel slits. Each slit produces its own two-dimensional spectral image on the detector, but offset from each other. Thus the spectral and spatial dimensions are mixed in the detector-X direction. The image separations are 0.39~\AA, corresponding to 4.5\as. Ordinarily the mixing of the spectral images would make the data difficult to analyze, but the narrow bandwidths of multilayer coatings are used to advantage to minimize the overlapping of the spectral images. By choosing relatively isolated lines in the spectrum at 108 (\ion{Fe}{xix}), 171 (\ion{Fe}{ix}) and 284~\AA\ (\ion{Fe}{xv}) it can be shown that plasma properties such as line-of-sight velocity, temperature and line broadening can be accurately extracted \citep{2019ApJ...882...13C}. The lines are formed at temperatures 10, 0.8 and 2.5~MK, respectively. Spatial resolution will be 0.4\as, comparable to the best yet obtained for an EUV instrument (the Hi-C rocket experiment), and spectral resolution will be 2500 to 5000, depending on the channel.

The key achievement of MUSE would be to effectively meet the goal of simultaneous two-dimensional imaging and high-resolution spectroscopy. Thus complex, highly dynamic features could be monitored on timescales of seconds without the highly-restricted view of a single-slit spectrometer. The downside is that the temperature coverage is not as complete as a single-slit spectrometer due to the limited number of channels. However, this is partly compensated for by an EUV multilayer context imager with two channels at 304 and 195~\AA, the former giving images of the upper chromosphere and the latter images at 1.5~MK, midway between the \ion{Fe}{ix} and \ion{Fe}{xv} lines.

\subsubsection{VERIS}

VERIS is a sounding rocket experiment led by the Naval Research Laboratory that has been funded for a second launch following the first in 2013. It is an imaging slit spectrometer covering the region 940--1140~\AA. The spectral range is comparable to the long-wavelength channel of SPICE, with a number of strong chromosphere, transition region and coronal emission lines. The  spectral and spatial  resolutions will be significantly better at around  5000 and 0.4\as, respectively.

\subsection{Slitless spectrometers}

\subsubsection{GIS on KORTES}\label{sect.gis}

KORTES (see article by Kirichenko et al.\ in the present issue) is an instrument package that will be installed on the ISS in 2024. In addition to X-ray instrumentation (Sect.~\ref{sect.crystal}) it will include a multilayer EUV imaging telescope, and two Grazing-Incidence Spectroheliographs (GIS). The latter are follow-ons to the TESIS \citep{2011SoSyR..45..162K} and SPIRIT \citep{2008AstL...34...33S} spectroheliographs that were flown on CORONAS-Photon and CORONAS-F, respectively.

Unlike slit spectrometers, the first optic in the path is a flat reflection grating which sends light in grazing incidence to a multilayer mirror and on to a CCD detector. The resulting image on the detector is a series of full-disk images, each corresponding to an emission line in the spectrum. A unique feature of the GIS design is that the grating is positioned such that the solar images are squashed by a factor of 20 in the dispersion direction, giving ``cigar"-shaped solar images instead of the circular disks shown in Fig.~\ref{fig.skylab}. This greatly reduces spectral/spatial confusion due to overlapping images. The spatial resolution in the dispersion direction is still sufficient to resolve compact features such as flares. 

One GIS on KORTES will operate in the 170--190~\AA\ and the other in the 280--335~\AA\ range. The emission lines in these ranges are mainly formed at temperatures 1--4~MK, but the strong upper chromosphere line \ion{He}{ii} 304~\AA\ will also be observed.

\subsubsection{COSIE}\label{sect.cosie}

COSIE is a novel EUV instrument concept from the Smithsonian Astrophysical Observatory that enables both wide-field EUV imaging of the solar disk and outer corona (to distances of 3~$R_\odot$), and slitless spectroscopy of the entire solar disk. A mechanism between the entrance aperture and the focus mirror has a flat mirror for the imaging mode, which can flip over to a grating for the spectroscopy mode. Both the mirror and grating have multilayer coatings that target the 186 to 205~\AA\ range. COSIE is not currently selected for any flight opportunities, but the concept along with the science justification is described in detail by \citet{2020JSWSC..10...37G}. Due to the slitless design, the spectral data would require deconvolution techniques to extract physical parameters from the images and this has been explored by \citet{2019ApJ...882...12W} using the method developed for MUSE (Sect.~\ref{sect.muse}).

\subsubsection{Hi-C/COOL-AID}\label{sect.cool-aid}

Hi-C is a sounding rocket EUV imaging instrument that was flown in 2012 \citep{2013Natur.493..501C} and 2018 \citep{2019SoPh..294..174R}, achieving the best spatial resolution of the corona yet obtained with a multilayer-based imaging telescope. (Note that IRIS has achieved a similar resolution when observing the long-wavelength \ion{Fe}{xxi} 1354.1~\AA\ emission line but does not use multilayer coatings.) The third flight is currently planned in 2024 as part of a special flare campaign whereby Hi-C and FOXSI (Sect.~\ref{sect.foxsi}) will be launched together while a flare occurs on the Sun. Hi-C will obtain images in a channel centered on the \ion{Fe}{xxi} 128.8~\AA\ line (formed at 11~MK). A new addition to Hi-C will be COOL-AID, a pair of identical slitless EUV spectrometers.

COOL-AID also targets the \ion{Fe}{xxi} 128.8~\AA\ line and the optical design is based on COSIE. Flares at this temperature are typically compact and Doppler shifts will lead to a smearing of the image in the solar-east direction, if blueshifts, and to solar-west if redshifts.
The second of the COOL-AID spectrometers is identical to the first, but the grating dispersion is in the orthogonal direction. Doppler shifts will thus lead to smearing in the north-south directions, rather than east-west. By combining the two images it will thus be possible to distinguish spatial structure from velocity structure.

This observation technique is a basic form of Computed Tomography Imaging Spectroscopy, which was recently investigated on the Sun with the ESIS sounding rocket experiment. Launched in 2019 and led by Montana State University, ESIS is a slitless spectrometer that feeds the solar image to four gratings, with dispersion directions that are progressively stepped by 45 degrees from each other. The wavelength band includes \ion{He}{i} \lam584.3 and \ion{O}{v} \lam629.7 (see Fig.~\ref{fig.skylab}). As with COOL-AID, Doppler shifts will result in smearing of compact bright points in the gratings' dispersion directions. Thus a small ``explosion" may result in smears in all of the gratings' images, but a collimated jet may only lead to smearing in one or two of the gratings' images.

\subsection{Non-imaging spectrometers}

\subsubsection{SunCET}\label{sect.suncet}

SunCET \citep{2021arXiv210109215M} is a CubeSat concept that is currently under Phase~A review with NASA. It has some similarities with COSIE in that it would have a compact EUV imager covering both the solar disk and the extended corona, with the aim of tracking coronal mass ejections. A wide-band multilayer coating will be used to cover the 170--200~\AA\ region. In terms of spectroscopy, there will be an EUV solar disk-averaged spectrometer operating from 170 to 340~\AA\ with modest 1~\AA\ resolution. The design is based on one of the EUV spectrograph channels of the EVE instrument on SDO \citep{2012SoPh..275..115W}.

\subsubsection{FURST}

The FURST rocket experiment, led by Montana State University, is scheduled for launch in 2022 and is unique amongst the EUV experiments considered here in covering a very broad wavelength range (1200--1800~\AA) with a high spectral resolution of 10\,000. Such a wide spectral range is usually prohibited by the limited size and/or number of CCD cameras for space instruments or, for EUV-A range instruments, the small bandpass of multilayer coatings. The optical design is also distinct with the reflecting surfaces being seven cylinders arranged along the Rowland circle \citep[see][for a description of a Rowland circle design applied to the UVCS instrument on SOHO]{1995SoPh..162..313K}. The cylindrical reflections result in the solar images being compressed to narrow slits in the direction parallel to the dispersion direction, thus giving narrow emission lines in the final spectrum despite the full-disk image information they contain. The mission goal is to obtain a Sun-as-a-star spectrum that can be compared with those obtained of other stars with the Hubble Space Telescope.



\section{The soft X-ray spectrum}\label{sect.sxr-spec}

At first glance the SXR spectrum seems less attractive for studying the solar atmosphere as there are no strong transition region lines apart from the \ion{C}{v} and \ion{N}{vi} lines at 40.7 and 29.1~\AA\ and, as discussed in Sect.~\ref{sect.instr}, there are difficulties in performing high resolution imaging spectroscopy. The coronal SXR lines are largely formed at temperatures of 5~MK or higher and so science objectives are usually focused on flares and active region heating. This has the consequence that any missions featuring SXR capability should ideally be flown prior to or during solar maximum. (The next maxima are expected to take place in the few years around 2025 and 2036.)

\begin{figure}[t]
    \centering
    \includegraphics[width=\hsize]{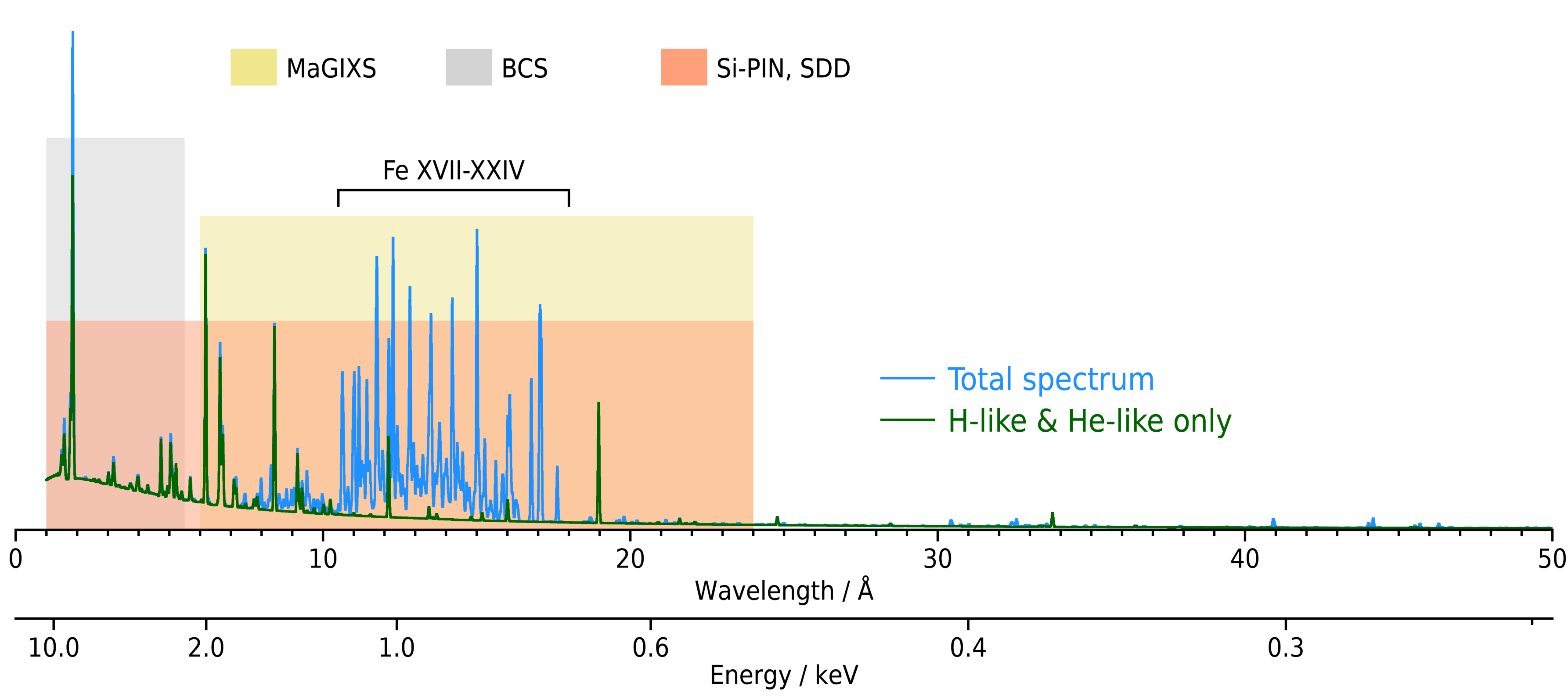}
    \caption{The blue line shows a synthetic flare spectrum generated with CHIANTI. The green line shows the same spectrum, but with only lines from H and He-like species. Boxes show the wavelength coverage of MaGIXS, bent crystal spectrometers (BCS), and Si-PIN/SDD detector spectrometers.}
    \label{fig.sxr}
\end{figure}

Figure~\ref{fig.sxr} shows a soft X-ray spectrum generated with CHIANTI over the region  1--50~\AA\ (0.1--12~keV). The three key features are (i) sets of H-like and He-like ions from carbon through nickel; (ii) lines from \ion{Fe}{xvii--xxiv} in the 11--17~\AA\ region; (iii) continuum emission that increases with energy.

The H and He-like ions of iron, \ion{Fe}{xxv} and \ion{Fe}{xxvi}, are of particular interest as they give rise to strong transitions at temperatures inaccessible to EUV wavelengths (\ion{Fe}{xxiv} is the hottest iron ion in the EUV, with a strong line at 192~\AA). Both species have been previously studied at high resolution with crystal spectrometers on board SMM and \emph{Yohkoh}.

Due to the wide range of elements available,  H and He-like ions also offer excellent FIP bias diagnostics, and examples include the analyzes presented in \citet{2008AdSpR..42..838S} and  \citet{2018ApJ...863...10P}. These diagnostics can also be applied to low-resolution spectra, as demonstrated by \citet{2015ApJ...802L...2C}.

The \ion{Fe}{xvii--xxiv} lines arise from transitions between $n=2$ and $n=3$ shells. (Weaker transitions between $n=2$ levels are found in the EUV between 90 and 140~\AA.) A large number of transitions are found between 11 and 17~\AA\ so high spectral resolution is required to resolve them sufficiently to apply the many temperature and density diagnostics that these ions offer. Combining the iron ions with the H-like ion \ion{Ne}{x} gives an excellent FIP bias diagnostic \citep[e.g.,][]{2001A&A...365L.324B}.

As noted in the Introduction, the continuum in the SXR region comes predominantly from free-free  and free-bound emission and is dominated by hydrogen, thus emission line to continuum comparisons can yield absolute element abundances.

\section{Soft X-ray instrumentation}\label{sect.sxr}

The only currently operating solar mission with SXR spectroscopic capability is STIX \citep{2020A&A...642A..15K} on Solar Orbiter, which was launched in 2020. It uses collimating grids (see Sect.~\ref{sect.imaging}) to enable a spatial resolution down to 7\as\ and, although mainly focused on hard X-rays, the instrument has coverage down to 4~keV but with a resolution of only 1~keV. There are also disk-averaged solar spectrometers currently flying on planetary missions as discussed in Sect.~\ref{sect.pin}. 

RHESSI, although principally a hard X-ray mission, had some SXR capability but ceased operations in 2018 after 16 years in orbit. The first MinXSS CubeSat was deployed from the ISS in 2016 and operated for one year, as planned. The follow-on MinXSS-2 \citep{2020AdSpR..66....3M} was intended to have a five-year lifetime after launch in 2018, but it failed within the first month.

\subsection{Silicon PIN and Drift detectors}\label{sect.pin}

Sect.~\ref{sect.detector} highlighted Si-PIN and SDD detectors for enabling very compact, non-imaging SXR spectrometers. The next evolution of the MinXSS concept will be the Dual-zone Aperture X-ray Solar Spectrometer (DAXSS), to be flown on INSPIRESat-1 in 2021. (INPSIRESat is a series of satellites developed by a consortium of space universities led by the University of Colorado at Boulder.) DAXSS was first flown on a sounding rocket \citep{2020ApJ...904...20S}. The key advance over the MinXSS instruments is an increase in the size of the primary aperture to increase sensitivity to high-energy photons, and the insertion of a Kaplon filter behind the primary to attenuate the more intense flux of low-energy photons. The result is a more balanced distribution of low and high-energy photons.

Another compact X-ray spectrometer utilizing SDDs will be SoLEXS \citep{2017CSci..113..625S}, which will be launched on the Indian Aditya-L1 spacecraft in 2022. Two SDDs will be used with apertures of different sizes in order to observe both large and small flares. Aditya-L1 will also carry HEL1OS, a disk-averaged hard X-ray spectrometer covering the 10--150~keV range.

\subsection{Crystal spectrometers}\label{sect.crystal}

Certain crystals can reflect X-rays at a specific wavelength, determined by the angle of incidence, as first determined by L.~and W.H.~Bragg in 1913. If the crystal is rotated to modify the incidence angle, then a spectrum can be built up. A single crystal will give access to a range $\approx\pm$40\%\ of the central wavelength. The Flat Crystal Spectrometer on SMM featured seven different crystals on a rotating drum structure, enabling almost complete wavelength coverage from 1.4 to 22.5~\AA. 

A problem with this design is that the position of a flare within the crystal spectrometer's field-of-view affects the position of the spectra on the detector, thus the shift of a spectral line could be due to plane-of-sky motions or Doppler motions. The ``dopplerometer" design resolves this ambiguity by using two identical crystals, but with opposite angles of incidence. Plane-of-sky motions will be in the same direction in both spectra, but Doppler motions will be in the opposite direction---see Figure~3 of \citet{2016ExA....41..327S}.

This approach was first attempted with the DIOGENESS instrument \citep{2015SoPh..290.3683S} on the Russian CORONAS-F mission. It had four separate channels, each covering a different wavelength region. Each of the dopplerometer channels had two crystals attached to a shaft that could rock backwards-and-forwards, allowing rapid scans through the spectral range. DIOGENESS was only operational for a few weeks in 2001 but some some flares were observed \citep{2015SoPh..290.3683S}.

The next iteration of this type of instrument will be part of the SOLPEX instrument package \citep{2014SPIE.9441E..0TS, 2019ExA....47..199S} to be flown with the Russian KORTES experiment to the ISS in 2024 (see article by Kirichenko et al.\ in this issue). The Rotating Drum X-ray Spectrometer (RDS) will have  crystals placed on an octagonal drum that rotates ten times per second to achieve high time resolution data of flares. High spectral resolution is enabled through the 1~$\mu$s readout time of SDDs. Radiation is incident on one of the outer halves of the drum, and detectors on the rotation and anti-rotation directions will enable dopplerometer spectra to be obtained.
The combination of crystals is chosen to  cover 0.4--23~\AA.

Bent crystal spectrometers (BCS) take advantage of Bragg diffraction to yield high resolution spectra over a narrow wavelength region without the need for scanning. They have been flown on SMM \citep{1981ApJ...244L.141C}, \emph{Yohkoh} \citep{1991SoPh..136...89C} and CORONAS-F \citep{2005SoPh..226...45S}. The ChemiX instrument \citep{2016ExA....41..327S} on Interhelioprobe will use bent crystals in two modes. One is a standard configuration with four spectral channels that combined give spectral coverage over 1.5--8.8~\AA. The other is a dopplerometer configuration with three spectral channels optimized to the H and He-like ions of iron, calcium and argon, respectively. Each channel has two crystals with angles of incidence in the opposite direction to yield the dopplerometer spectra. Because bent crystals are used, there is no need to rotate the crystals.

\subsection{MaGIXS}

Imaging slit spectrometers have been very successful at EUV wavelengths, but none have been flown in the SXR range. This will change with the launch of the sounding rocket experiment MaGIXS \citep{2018SPIE10699E..27K} in 2021. It will observe the 6--24~\AA\ region without the need for scanning and have spatial resolution along the slit of 5\as. The spectral resolution of 500 is not as good as that possible with the crystal spectrometers, but significantly better than for the silicon detector instruments. 

Simultaneous coverage of the \ion{Fe}{xvii--xxiii} lines with the H and He-like lines of oxygen through silicon will lead to excellent plasma diagnostics in the 3--10~MK temperature range. This is important for understanding the heating of solar active regions. The spectra would also be excellent for flare measurements, but this is unlikely for a 5-minute rocket flight.

\subsection{FOXSI}\label{sect.foxsi}

FOXSI is a sounding rocket experiment that was flown in 2012, 2014 and 2018. It was the first solar instrument to obtain solar images at hard X-ray wavelengths using focusing optics \citep{2014ApJ...793L..32K}, and the detectors yield modest spectral resolution. Wavelength coverage extends to the 4--10~keV SXR region, where spectral resolution was around 10--20. The data have been valuable for studying active region heating \citep{2017NatAs...1..771I, 2020ApJ...891...78A}.

The FOXSI concept was proposed for the 2016 SMEX call, but was not selected after reaching Phase~A.  For the 2019 MIDEX call it was teamed with an EUV imager and the concept was named FIERCE, but it was not selected for Phase~A. A key difference with the rocket experiments is that the spacecraft versions would have had extendable booms, extending the spectral coverage from around 20~keV up to around 70~keV.

The next FOXSI flight is planned for 2024 as part of a special flare campaign coordinated with another launch of the Hi-C EUV imager. The latter will include the COOL-AID EUV spectrometer (Sect.~\ref{sect.cool-aid}).

\subsection{CubIXSS}

CubIXSS is a CubeSat mission concept that is currently under Phase~A review with NASA. It consists of two distinct instruments: a spatially-integrated SXR spectrometer, the Small Assembly for Solar Spectroscopy (SASS), that is an evolution of the MinXSS spectrometers, and the Multi-Order X-ray Spectral Imager (MOXSI). The latter has a set of five pinhole cameras, four of which yield full-disk images of the Sun in X-ray lines. The fifth camera feeds a transmission grating to yield full-disk, overlappogram spectra (see Fig.~\ref{fig.skylab}). Further details are available in the White Paper of \citet{2017arXiv170100619C}.

\subsection{Mission concepts}

\subsubsection{PhoENIX}\label{sect.phoenix}

Physics of Energetic and Non-thermal Plasmas in the X-region (PhoENIX) is a Japanese mission concept currently considered for the solar maximum in the mid-2030's. It consists of three separate instruments: Soft X-ray Imaging Spectrometer (SXIS), Hard X-ray Imaging Spectrometer (HXIS) and Soft Gamma-Ray SpectroPolarimeter (SGSP). 

SXIS would combine a Wolter-type telescope with spatial resolution around 1\as\ with a photon-counting CMOS sensor. The latter yields high time resolution and modest spectral resolution (comparable to the silicon drift detectors), and was first flown on the FOXSI-3 sounding rocket in 2018 \citep{2019SpReT.205...27.}.

\subsubsection{SSAXI}\label{sect.ssaxi}

The SmallSat Solar Axion and Activity X-ray Imager (SSAXI) is a SmallSat concept proposed by the Smithsonian Astrophysical Observatory \citep{2019SPIE11118E..10H} that would perform imaging spectroscopy in the 0.6--6~keV range with spatial and spectral resolutions of 30\as\ and 10, respectively. There are six small Wolter-type telescopes feeding CMOS sensors with energy-resolving capability (similar to that used on the FOXSI-3 flight). Unlike other X-ray missions, SSAXI would be focused on small-scale flaring activity in the quiet Sun and active regions and so would be preferentially launched during solar minimum.

\section{Summary}\label{sect.summary}

This article has attempted to identify the new technologies and missions that will operate at soft X-ray and extreme ultraviolet wavelengths in the years to come. These wavelength ranges are critical for understanding energy and mass flow through the solar atmosphere and how million-degree plasma is heated and maintained. The missions are listed in Table~\ref{tbl.summary} and classified according to whether they are on the way, are proposed, or whether they are just concepts at this point.

I note that there are currently no large STP or LWS-class solar missions planned for the future and so opportunities for the traditional large, high-resolution instruments flown in the past are limited to smaller Explorer-class missions and missions-of-opportunity. However, as hopefully demonstrated in this article, there are many innovative, low-cost CubeSat and SmallSat concepts that should yield exciting new results in the future.

For further reading, the introduction of \citet{2016JGRA..121.8250C} provides a good summary of the types of spectrometers used in Solar Physics. \citet{2016JGRA..12111667H} discusses SXR and EUV instrument options for investigating magnetic energy release in the solar atmosphere, particularly in relation to solar flares. The \emph{Next Generation Solar Physics Mission Report}\footnote{\url{https://hinode.nao.ac.jp/SOLAR-C/SOLAR-C/Documents/NGSPM_report_170731.pdf}.} was commissioned by JAXA, NASA and ESA and written by a panel of Solar Physics experts based on community input. It recommended three instrument types for studying fundamental physical processes in the solar atmosphere at high resolution. The report gives a valuable discussion of the important scientific issues and instrumentation options. The selection of Solar-C\_EUVST (Sect.~\ref{sect.euvst}) satisfies one aspect of this recommendation.

\begin{table}[t]
\caption{Summary of future EUV and SXR spectroscopy missions.}
\begin{center}
\begin{tabular}{llll}
\noalign{\hrule}
\noalign{\smallskip}
\noalign{\hrule}
\noalign{\smallskip}
& On the way & Proposed & Long term\\
\noalign{\hrule}
\noalign{\smallskip}
EUV & EUNIS & MUSE & COSIE\\
    & Solar-C\_EUVST & SunCET \\
    & FURST &\\
    & VERIS \\
    & Hi-C/COOL-AID \\
    & SNIFS\\
     & KORTES/GIS \\
\noalign{\medskip}
SXR & MaGIXS & CubIXSS & PhoENIX  \\
    &INSPIRESat-1/DAXSS & & SSAXI\\
    & KORTES/SOLPEX & & Microcalorimeters\\
    & FOXSI-4 & & \\
    & Aditya-L1/SoLEXS \\
    & Interhelioprobe/ChemiX \\
\noalign{\hrule}
\end{tabular}
\end{center}
\label{tbl.summary}
\end{table}

\section*{Conflict of Interest Statement}

The author declares that the research was conducted in the absence of any commercial or financial relationships that could be construed as a potential conflict of interest.

\section*{Author Contributions}

All text and figures were created by the author.

\section*{Funding}
Funding was provided by the \emph{Hinode} project and the NASA Heliophysics Guest Investigators program.

\section*{Acknowledgments}
I thank the three anonymous referees for their valuable comments. I also thank  Amy Winebarger, Adrian Daw, Sabrina Savage, Edward DeLuca and Phil Chamberlin for giving information about some of the projects discussed in this work.

\bibliographystyle{frontiersinSCNS_ENG_HUMS} 
\bibliography{test}


\end{document}